\documentclass[nobibnotes,aps,superscriptaddress,10pt,notitlepage,longbibliography,nofootinbib]{revtex4-1}
\usepackage[utf8]{inputenc}
\usepackage[T1]{fontenc}
\usepackage{comment}

\usepackage{graphicx}
\graphicspath{{Figures/}}
\usepackage{float}
\usepackage{latexsym}
\usepackage{graphicx}
\usepackage{amssymb}
\usepackage{booktabs}
\usepackage{amsmath}
\usepackage{amsfonts}
\usepackage{physics}
\usepackage[colorlinks=true,citecolor=blue,hyperfootnotes=false]{hyperref}
\usepackage[dvipsnames]{xcolor}
\usepackage{dcolumn}
\usepackage{textcomp}
\usepackage{xcolor}
\usepackage{xfrac}
\usepackage{slashed}
\usepackage{multirow}
\usepackage[scr=esstix]{mathalpha}

\usepackage{xcolor}

\begin{document}

\title{Seesaw and Axion in No-Scale Gravity}

\author{Anamaria Hell}
\affiliation{Center for Data-Driven Discovery,\\ Kavli Institute for the Physics and Mathematics of the Universe (WPI),
UTIAS, The University of Tokyo, Chiba 277-8583, Japan}

\author{Lincoln da S. Pereira}
\affiliation{Department of Mathematical Physics,
Institute of Physics, University of Sao Paulo, R. do Matao 1371, Sao Paulo, SP 05508-090, Brazil}

\author{Tsutomu T. Yanagida}
\affiliation{Kavli Institute for the Physics and Mathematics of the Universe (WPI),
UTIAS, The University of Tokyo, Chiba 277-8583, Japan}
\affiliation{Tsung-Dao Lee Institute \& School of Physics and Astronomy, Shanghai Jiao Tong University, Pudong New Area, Shanghai 201210, China}

\begin{abstract}
{The No-Scale gravity is a compelling framework to describe particle physics, gravity and cosmology, in which all mass scales are an illusion given by the value of a scalar field $\phi$. This theory, however, falls under the umbrella of more general scalar-tensor theories, which propagate different modes depending on the scalar field, and faces the extensive debate regarding the equivalence between its Jordan and Einstein frame descriptions. In this work we advocate that by excluding the singular point in the transformation between the two frames (e.g., $\phi=0$) we can have the classical equivalence between them.  
To show the merit of removing this singular point, we argue the Jordan frame makes the symmetries of the model manifest, which creates a suitable playground for the particle-physics model building which we illustrate by showing the emergence of discrete symmetries which leads to a high-quality axion and also naturally incorporate the seesaw mechanism.}

\end{abstract}

\maketitle

\tableofcontents

\section{Introduction}\label{sec:intro}

No-Scale Gravity \cite{Hong:2025tyi,Hong:2025cae} is a  special case of scalar-tensor theories, that describes a scalar field with a kinetic term and non-minimal coupling to the Ricci scalar but no Einstein term. Inspired by the Brans-Dicke gravity \cite{Brans:1961sx}, it falls into a class of more general Variable gravity theories \cite{Wetterich:1987fk, Wetterich:1987fm, Wetterich:2013jsa}, and theories with non-minimally coupled curvature with power laws \cite{Hell:2025lbl}\footnote{ 
The theory also closely resembles the so called \textit{induced gravity theories} \cite{Zee:1979hy,Rinaldi:2015uvu,Cooper:1981byv}. However, while the gravitational sector of both theories are similar, No-Scale Gravity includes a scale-invariant version of the Standard Model (SM)
which is usually not dealt with in the induced gravity theories. Due to the way No-Scale Gravity introduces the SM sector, there are no new long range forces. Also, another difference is that it does not rely on the assumption of dynamical spontaneous symmetry breaking , which is something ubiquitous in induced gravity theories.}.

The theory discussed here is an extension which is particularly interesting, since it does not only replace the Planck mass, but also all other dimensionful parameters of the Standard Model by a scalar $\phi$, hence the name No-Scale gravity. Thus, the theory is perfectly scale-invariant \cite{Wetterich:2019qzx} and the Planck mass $M_P$ is an illusory parameter. This theory has a completely massless dilaton and the exchanges of the dilaton never induce any long range forces because it interacts with matter always with its derivative couplings \cite{Hong:2025tyi}. This idea has shown to be promising, as in the previous work No-Scale gravity seems to be able to accommodate candidates for dark energy and dark matter \cite{Pereira:2026llu}. In addition, it is also compatible with Starobinsky inflation \cite{Hong:2025tyi}, providing in this way a seemingly complete picture, though further work is necessary.

Originally, the No-Scale Gravity is formulated in the Jordan frame, in which its scale invariance is manifest. At the same time, its formulation in the Einstein frame might sometimes be more appealing. In it, the gravitational contribution takes the form of the Einstein term, coupled to the scalar field with a more complicated potential.
However, as it was recently pointed out in \cite{Hell:2025lbl}, theories with non-minimally coupled curvature may have special points in the Jordan frame depending on the space-time background, and the background values of the scalar field. This dependence manifests through the different number of propagating modes. One important consequence of this change is that  Einstein and Jordan frames do not necessarily describe the same physics for all values of the metric and the scalar field -- the points at which the number of degrees of freedom changes also often correspond to the points at which the transformation between the two frames is singular.

One simpler example of this is the \textit{pure} $R^2$ theory, which has no propagating modes for Minkowski space-time, and Schwarzschild and Kerr black holes, while at the same time propagating three degrees of freedom for a general background in the Jordan frame\footnote{The same result can be also extended to pure scale-invariant action in d-dimensions \cite{Hell:2025wha, Hell:2026pwm}.} \cite{Hell:2023mph, Barker:2025gon}.  However, when written in the Einstein frame, the theory takes the form of a Ricci scalar together with a scalar field with no potential term and a cosmological constant, losing the information about background solutions which contain no propagating modes in the Jordan frame. The main reason for Einstein frame not being able to accommodate the full content of $R^2$ gravity in the Jordan frame is because the transformation between the two is singular for space-times with vanishing Ricci scalar, breaking the condition for the equivalence between the two frames for the $f(R)$ theories of gravity \cite{Sotiriou:2008rp}.

While No-Scale Gravity is indeed an intriguing theory for the standard model in particle physics, transforming it from Jordan to Einstein frame also contains singular points, meaning that the two are not fully equivalent. In particular, this takes place when the Brans-Dicke type scalar vanishes ($\phi=0$). As we will see, this is a point where the physics of No-Scale Gravity is also unclear in the Jordan frame, which further motivates its removal\footnote{We will investigate in future work the theory at $\phi=0$, as well as the effects of the quantization of the theory.}.  Fortunately, the $\phi=0$ point is a fixed point of the scale transformation, and thus we can remove it safely from the theory. In this work we will consider a special case of No-Scale gravity theory which does not contain this  origin of field space, more explicitly $\phi\in \mathbb{R}/\{0\}$.


Given this, we wish to push forward that one should consider the Jordan frame as more suitable playground for model building, by relying on the classical Einstein and Jordan frame equivalence, restored in this particular formulation of No-Scale gravity. The key motivation for this is the manifestation of symmetries which are otherwise hidden in the Einstein frame. In this letter, we will illustrate this case explicitly by considering a discrete gauge symmetry $\mathbb{Z}_4\times \mathbb{Z}_3\times \mathbb{Z}_5$ inspired by previous work \cite{Sheng:2025sou}, where both the fermions and BD scalar are charged under it and we will be able to naturally incorporate the seesaw mechanism. Also by a little extension of the SM we will be able to accommodate a high-quality axion field that solves the strong CP problem \cite{Peccei:1977ur,Weinberg:1977ma,Wilczek:1977pj}. 

As a result, in this letter, we will show that the seesaw mechanism \cite{Minkowski:1977sc, Yanagida:1979as, Ramond:1979ijt} \footnote{For further extension see \cite{Yanagida:1979gs,Gell-Mann:1979vob,Yanagida:1980xy}.} arises naturally in the No-Scale Gravity and a little extension of the SM sector accommodates the high quality QCD axion almost automatically.

\section{No-Scale Gravity without $\phi=0$ }\label{sec:model}
As indicated in the introduction, the No-Scale gravity theory \cite{Hong:2025tyi} is based on the idea that not only the Planck scale, but all scales are an illusion.  All dimensionfull parameters are replaced by the BD scalar $\phi$, such that the Lagrangian density may be written as \footnote{Here we use the $(+,-,-,-)$ signature and define the Ricci tensor as $R_{\mu\nu}=-R^{\rho}_{\;\mu\rho\nu}.$}

\begin{align}\label{eq:lagrangian_jordan}
    \frac{\mathcal{L}}{\sqrt{-g}} = \frac{1}{2}\xi\phi^2 R + \frac{1}{2}\partial_\mu\phi\partial^\mu \phi - \frac{\lambda}{4}\phi^4 + \frac{\mathcal{L}_{\rm SM}}{\sqrt{-g}},
\end{align}
where $\mathcal{L}_{SM}$ denotes the Lagrangian for matter, involving the Standard Model and also the scalar $\phi$ coupled in such a way that the global scale symmetry is preserved.
While this theory has originally been defined with scalar field taking all possible values in the field space, we will now investigate the possibility to remove the $\phi=0$ point. There are several reasons to consider this possibility. First, we can notice that the equations of motion do not uniquely determine the background evolution\footnote{The background might be instead determined by the higher order perturbations, as we will discuss in a future work.}. To see this, we note that the equations of motion for the metric, and the scalar field in the absence of $\mathcal{L}_{SM}$ are respectively given by: 
\begin{equation}
    \xi \phi^2R_{\mu\nu}-g_{\mu\nu}\left(\frac{1}{2}\xi \phi^2 R+\partial_{\gamma}\phi\partial^{\gamma}\phi-\frac{\lambda}{4}\phi^4\right)+\partial_{\mu}\phi\partial_{\nu}\phi+\xi\left(\nabla_{\mu}\nabla_{\nu}\phi^2-g_{\mu\nu}\nabla_{\gamma}\nabla^{\gamma}\phi^2\right)=0
\end{equation}
and
\begin{equation}
    \nabla_{\mu}\nabla^{\mu}\phi-\xi R\phi+\lambda \phi^3=0. 
\end{equation}
We can notice that for $\phi=0$, both equations are automatically satisfied, making this choice a solution of the theory. However, at the same time, the curvature is completely arbitrary at this order, and might thus be instead determined by the higher-order perturbations. 
Moreover, the number of propagating modes will change depending on the contribution of $\phi$. In order to illustrate this, let us consider the perturbations around metric and the scalar field:
\begin{equation}
    g_{\mu\nu}=g_{\mu\nu}^{(0)}+\delta g_{\mu\nu},\qquad \text{and}\qquad \phi=\phi^{(0)}+\delta \phi, 
\end{equation}
where $g_{\mu\nu}^{(0)}$ and $\phi^{(0)}$ denote the background values, and set Friedmann–Lemaître–Robertson–Walker background, with a spatially flat metric
\begin{equation}
    ds^2=dt^2-a^2\delta_{ij}dx^idx^j, 
\end{equation}
and $\phi^{(0)}=0$. By decomposing the metric perturbations according to the group of spatial rotations, 
\begin{equation}
    \begin{split}
        \delta g_{00}&=2\Phi\\
        \delta g_{0i}&=S_i+\partial_i B\\
        \delta g_{ij}&=a^2\left(2\Psi\delta_{ij}+2\partial_i\partial_j E+\partial_i F_j+\partial_j F_i + h_{ij}^T\right),
    \end{split}
\end{equation}
where
\begin{equation}
    \partial_i S_i=0,\qquad \partial_i F_i=0,\qquad h^T_{ii}=0,\qquad \text{and},\qquad \partial_i h_{ij}^T=0, 
\end{equation}
we find the Lagrangian density at the quartic order in the perturbations that involves only the scalar field perturbations: 
\begin{equation}\label{secondorderL}
    \mathcal{L}=\frac{a^3}{2}(\delta\dot{\phi})^2-\frac{1}{2}a\left(k^2-6\xi\left(\Ddot{a} a + \dot{a}^2\right) \right)\delta\phi^2,
\end{equation}
where dot denotes the derivative with respect to the coordinate time $t$. At the same time, if $\phi$ starts to contribute to the background, the number of propagating modes is going to change, giving rise to the gravitational waves. This is easy to see from the first term in (\ref{eq:lagrangian_jordan}) which becomes an effective Planck mass multiplying the Ricci scalar. 
It should be also noted that the scale-factor $a$ in (\ref{secondorderL}) is arbitrary -- it cannot be fixed by the background equations of motion alone\footnote{Although, one can consider higher-order contributions, which we will discuss in a companion paper. }. Due to this degeneracy, as well as the change in the number of propagating modes, the case with $\phi=0$ is a special point. This also justifies our aim to make the No-Scale Gravity without the origin of $\phi$ self-consistent.

This previous degeneracy in the arbitrariness of the background can also be relaxed by the addition of the $R^2$ term to the action\footnote{The inclusion of the $R^2$ term is very important for generation of the Starobinsky inflation. We will discuss details about the inclusion of this term in future publications.}(\ref{eq:lagrangian_jordan}):
\begin{align}\label{eq:lagrangian_jordan2}
    \frac{\mathcal{L}}{\sqrt{-g}} = \frac{1}{2}\xi\phi^2 R +\beta R^2 + \frac{1}{2}\partial_\mu\phi\partial^\mu \phi - \frac{\lambda}{4}\phi^4  + \frac{\mathcal{L}_{\rm SM}}{\sqrt{-g}}
\end{align}
which is also scale-invariant.
However, both theories will not be fully equivalent to the Einstein frame formulation due to the singular points in the Weyl transformations, necessary to relate the Jordan and Einstein frames. In the first case, with Lagrangian density (\ref{eq:lagrangian_jordan}), to reach the theory in the Einstein frame, one needs to perform the Weyl  transformation:  
\begin{equation}
     g^{E}_{\mu\nu}= \Omega g^{J}_{\mu\nu},\qquad  \text{where}\qquad \Omega^2 = \xi\phi^2/M^2_p,
\end{equation}
and $ g^{E}_{\mu\nu}$ and $ g^{J}_{\mu\nu}$ denote the metric in Einstein and Jordan frame respectively. Clearly, this transformation is singular for $\phi=0$. In the more general case (\ref{eq:lagrangian_jordan2}) one should first introduce an additional scalar field in order to reduce the power of the Ricci scalar. However, then one comes to the same conclusion -- the transformation between Einstein and Jordan frames will not be regular for all values of the scalar, and the two frames lead to different number of propagating modes in backgrounds such as the flat space-time \cite{Hell:2025lbl}. Therefore, without the loss of generality, in the following, we will focus on the case described by the Lagrangian density (\ref{eq:lagrangian_jordan}), and consider the possibility to exclude the $\phi=0$ point.

Now we devote some attention to one of the major points of No-Scale gravity, which is the SM part in (\ref{eq:lagrangian_jordan}). Following the idea of keeping scale invariance, the no-scale gravity theory includes the Higgs ($H$) mass term and the right-handed neutrino ($N$) mass 
\begin{align}\label{eq:coupling}
    \mathcal{L}_{\rm SM}\supset \lambda_h\phi^2|H|^2+\frac{1}{2}\lambda_N \phi \bar{N}\bar{N}.
\end{align}
This theory is exactly scale-invariant at the classical level \cite{Wetterich:2019qzx}. Due to this we don't have the presence of long-range forces. Notice that all terms that have only dimensionless coupling constants are the same as those in the SM. This invariance is also kept at the quantum level by also replacing the cut-off scale by $\phi$, which is referred as \textit{scale-invariant renormalization scheme} \cite{Englert:1976ep, Shaposhnikov:2008xi,Armillis:2013wya,Hamada:2016onh,Falls:2018olk}.\\

As indicated before we chose to remove the $\phi=0$ point. However, such a removal renders the $\phi$ domain disconnected, meaning that $\phi:\mathbb{R}^n\to (-\infty,0)\cup (0,\infty) $. To connect the two domains, by looking the gravitational part of Lagrangian (\ref{eq:lagrangian_jordan}) we can notice a discrete symmetry $ \mathbb{Z}_2$ which would help to make the domain simply connected. Motivated by the internal structure of the SM we enlarge such a symmetry to a $\mathbb{Z}_4$ gauge symmetry. However, the $\mathbb{Z}_4$ is anomalous, it has the Dai-Freed (DF) anomalies \cite{Dai:1994kq,Yonekura:2016wuc,Garcia-Etxebarria:2018ajm}. In order to cure this we introduce the standard-model quarks and leptons \cite{Kawasaki:2023mjm} and consider the $\mathbb{Z}_4$ as a gauge symmetry. We give the corresponding charges as in Table \ref{tab:Z4charge}, where the $T$ and $\bar F$ are $\boldsymbol{10}$ and $\boldsymbol{5^*}$ of the $SU(5)$. \footnote{We use the notation of the $SU(5)_{GUT}$ for a notation convenience only. Thus, we do not unify the SM gauge groups.} Hence the full Lagrangian is clearly invariant under such a discrete symmetry, which is anomaly free and allow us to have $\phi:\mathbb{R}^n \to  (0,\infty)$ which is simply connected. \\

\begin{table}[h]
    \centering
    \setlength{\tabcolsep}{12pt}
    \begin{tabular}{cccccc}
        \hline\\[-2.0ex]
         & $T$ & $\bar{F}$ & $\bar{N}$ & $\phi$ & H \\
        \hline\hline
        $\mathbb{Z}_4$ & 1 & 1 & 1 & 2 &  2 \\ 
        \hline
    \end{tabular}
    \caption{Charges of the fermions, the BD scalar ($\phi$) and the Higgs bosons ($H$) under $\mathbb{Z}_4$.}
    \label{tab:Z4charge}
\end{table}

The $\mathbb{Z}_4$ charge of the $\phi$ is 2 and hence we have effectively only its $\mathbb{Z}_2$ subgroup. The Higgs $H_{1,2}$ have also the $\mathbb{Z}_4$ charge 2 and hence they have trivial charges of the subgroup $\mathbb{Z}_2$. This is the reason why the Higgs condensations never produce domain walls. The same reason can be applied for the quark-quark condensation in QCD. In addition, not only the scale invariance, but also the $\mathbb{Z}_4$ gauge symmetry are only manifest in the Jordan frame Lagrangian. For this reason, we propose to identify the Jordan frame with $\phi\neq 0$ as the suitable frame for studying the symmetries.

\section{Discrete Gauge Symmetries}\label{sec:Z4}

We now extend the previous model to include three families of quarks and leptons, $T_i, \bar F_i, \bar N_i$ for $(i=1,...,3)$ and impose an anomaly free $\mathbb{Z}_3$ gauge symmetry. Previously \cite{Sheng:2025sou}, it was needed to introduce three Weyl fermions $\chi_i$ to cancel the Dai-Freed anomaly of the $\mathbb{Z}_3$ \cite{Dai:1994kq,Yonekura:2016wuc,Garcia-Etxebarria:2018ajm}. However, here as is shown in appendix \ref{app:Dai-Freed_Anomaly}, we can avoid the anomaly without introducing the extra three fermions in the spectrum. Still we need two Higgs doublets\footnote{Note that if we have a single Higgs we could have $TTH$ invariant under $\mathbb{Z}_3$ but then $T\bar{F}H^{\dagger}$ won't.}, $H_1, H_2$ to give masses for all quarks and leptons. The Yukawa coupling terms are given by
\begin{align}\label{eq:yukawa coupling}
    \mathcal{L}_{\rm SM}\supset y_{i,j} T_iT_jH_1+ y'_{i,j}T_i\bar F_j H_2^\dagger + y''_{i,j} \bar F_i\bar N_jH_2.
\end{align}
Here, we have automatically a global $U(1)$ symmetry that is the independent $U(1)$ rotation of the $H_1$ and $H_2$. This is nothing but the Peccei-Quinn (PQ) symmetry. It should be pointed out that the discrete gauge symmetry $\mathbb{Z}_3$ (and equivalently the three families of the quarks and leptons) is crucial for the presence of the global PQ symmetry.
Further, we should introduce a new singlet scalar field $\Phi_{PQ}$ to induce a spontaneous breaking of the PQ symmetry. The $\Phi_{PQ}$ has a coupling as
\begin{align}\label{eq:pq-coupling}
    \mathcal{L}_{\rm SM}\supset \lambda_{PQ} \Phi_{PQ}^2 H_1^\dagger H_2+ h.c. 
\end{align}
Notice that the above coupling supplies the mixing mass for the two Higgs doublets $H_1$ and $H_2$ that is necessary for the consistent phenomenology. The vacuum expectation value of the $\Phi_{PQ}$ generates the spontaneous breaking of the global $PQ$ symmetry and a massless Nambu-Goldstone boson called the axion $a(x)$. The charge of the fields under this $\mathbb{Z}_3$ is given in Table \ref{tab:Z4xZ3}

\begin{table}[h]
    \centering
    \setlength{\tabcolsep}{12pt}
    \begin{tabular}{ccccccccc}
        \hline\\[-2.0ex]
         & $T$ & $\bar{F}$ & $\bar{N}$ & $\phi$ & $H_1$ & $H_2$ & $\Phi_{\rm PQ}$  \\
        \hline\hline
        $\mathbb{Z}_4$ & 1 & 1 & 1 & 2 &  2 & 2 & 2  \\ 
        \hline
        $\mathbb{Z}_3$ & 1 & 1 & 0 & 0 &  -2 & 2 & 1 \\ 
        \hline
    \end{tabular}
    \caption{Charges of the fermions, the BD scalar ($\phi$), the Higgs bosons ($H$) and the PQ field $\Phi$ under $\mathbb{Z}_4\times \mathbb{Z}_3$.}
    \label{tab:Z4xZ3}
\end{table}

However, this model does not succeed in solving the strong CP problem. There is a dimension 6 $\mathbb{Z}_4\times \mathbb{Z}_3$ gauge invariant operator
\begin{align}\label{eq:coupling}
    \sim \frac{\Phi_{PQ}^6}{\xi\phi^2},
\end{align}
which makes the axion too heavy for the PQ mechanism to work.

Fortunately, we can introduce another independent anomaly free discrete gauge symmetry $\mathbb{Z}_5$ in the above framework. We show all charges of the discrete symmetries in Table \ref{tab:Z4xZ3xZ5}. For this choice of charges, we see that the symmetry $\mathbb{Z}_5$ is the unique choice
\footnote{The uniqueness of $\mathbb{Z}_5$ follows by looking at Table \ref{tab:Zn_non_anomalous}, we wish to avoid giving any extra charge to $\phi$ hence because of the $\phi \bar{N}\bar{N}$ coupling we may also take $N$ with 0 charge. This already singles out the $\mathbb{Z}_5$ as the only anomaly-free option.}
of extra non-anomalous symmetry we can introduce. Now the mixing for the $H_1$ and $H_2$ is given by
\begin{align}\label{eq:coupling}
   \sim\lambda_{PQ}\frac{H_1^\dagger H_2}{(\sqrt{\xi}\phi)^3}\Phi_{\rm PQ}^5.
\end{align}

\begin{table}[h]
    \centering
    \setlength{\tabcolsep}{12pt}
    \begin{tabular}{ccccccccc}
        \hline\\[-2.0ex]
         & $T$ & $\bar{F}$ & $\bar{N}$ & $\phi$ & $H_1$ & $H_2$ & $\Phi_{\rm PQ}$ \\
        \hline\hline
        $\mathbb{Z}_4$ & 1 & 1 & 1 & 2 &  2 & 2 & 2  \\ 
        \hline
        $\mathbb{Z}_3$ & 1 & 1 & 0 & 0 &  -2 & 2 & 1 \\ 
        \hline
        $\mathbb{Z}_5$ & 1 & 2 & 0 & 0 &  -2 & -2 & 1 \\ 
        \hline
        $U(1)_{PQ}$ & -2 & 1 & 0 & 0 &  4& -1 & 1\\ 
        \hline
    \end{tabular}
    \caption{Charges of the fermions, the BD scalar ($\phi$), the Higgs bosons ($H$) and the PQ field $\Phi$ under $\mathbb{Z}_4\times \mathbb{Z}_3\times \mathbb{Z}_5\times U(1)_{\rm PQ}$.}
    \label{tab:Z4xZ3xZ5}
\end{table}

Remarkably, the lowest higher dimensional relevant operator which is allowed by the gauge symmetries, but that violates the PQ symmetry, is \footnote{There are in fact lower dimensional operators, $(H_1^\dagger H_2)^4 \Phi_{PQ}^5/\phi^9$ and $(H_1H_2^\dagger)\Phi_{PQ}^{10}/\phi^8$, which are negligible. The contribution to the mass would be smaller compared to (\ref{eq:PQbreaking}) since the VEV of $H_i$ are much smaller than the PQ breaking scale.}
\begin{align}\label{eq:PQbreaking}
   \sim g\frac{\Phi_{PQ}^{15}}{(\sqrt{\xi}\phi)^{11}}.
\end{align}
We can easily confirm the induced mass for the axion is sufficiently suppressed and the strong CP problem is solved by the PQ mechanism. Here we show the piece of the Lagrangian in the Jordan frame, with the family indexes suppressed, which is relevant for the present discussion,
\begin{align}\label{eq:Jordan-frame}
    \mathcal{L}_J&\supset \frac{1}{2}\xi\phi^2R+ \frac{1}{2}\partial_\mu \phi \partial^\mu \phi - \frac{\lambda}{4!}\phi^4+  \frac{1}{2}\lambda_N\phi \bar{N}\bar{N} + \frac{1}{2}y TTH_1 +y'T\bar{F}H_2^{\dagger} + y''\bar{N}\bar{F}H_2  +h.c. \nonumber\\
     &+ \lambda_{\rm PQ}\frac{H^\dagger_1 H_2\Phi_{\rm PQ}^5}{(\sqrt{\xi}\phi)^3} + g\frac{\Phi_{PQ}^{15}}{(\sqrt{\xi}\phi)^{11}}+ \frac{1}{2}\partial_\mu \Phi_{\rm PQ}\partial^\mu \Phi_{\rm PQ}^{\dag}+...
\end{align}

It is now clear that we have a global symmetry $U(1)_{PQ}$, which is explicitly broken only by the higher dimensional operator $g\frac{\Phi_{PQ}^{15}}{M_P^{11}}$ in (\ref{eq:PQbreaking}). For the charge assignment of the global $U(1)$ symmetry for each field see Table \ref{tab:Z4xZ3xZ5}. Moreover, the inverse of the field further reinforces the choice of excluding the vanishing values of $\phi$, which would otherwise give rise to divergences. 


\section{Seesaw, Axion and Their Phenomenology}

Although we discussed the symmetries in the Jordan frame, we move to the Einstein frame to discuss the physics, which follows by taking the Weyl  transformation $\Omega^2 = \xi\phi^2/M^2_p$ and $g_E= \Omega g_J$, where the subindex denotes the Einstein frame ($g_E$) and Jordan frame ($g_J$) metric. We drop the subindex for simplicity. The change of frames need some care when dealing with scalar fields, the fermion and gauge fields are all Weyl invariant. However, the scalar fields become non-canonically normalized after changing frame, that is why we take $\Phi_{\rm PQ}\to \Omega \Phi_{\rm PQ}$ and $H\to \Omega H$ which makes it explicit the derivative coupling between the dilaton ($\chi = M_p \sqrt{(1+\frac{6}{\xi})}\ln\Omega$)\footnote{In the Einstein frame we have $-\infty<\chi<\infty$ which corresponds to $\phi \in (0,+\infty)$ in the Jordan frame. This is consistent since we removed $\phi=0$ and have a $\mathbb{Z}_4$ symmetry.} and these scalars \cite{Hong:2025tyi}. However, we omit the derivative coupling. Hence, the Lagrangian in the Einstein frame, with the family indexes suppressed, becomes 
\begin{align}\label{eq:Einstein-frame}
    \mathcal{L}_E&\supset \frac{1}{2}M_{\rm P}^2R+ \frac{1}{2}\partial_\mu \chi \partial^\mu \chi - \frac{\lambda M_{\rm P}^4}{4!\xi}+  \frac{1}{2}\lambda_{N}M_{\rm P} \bar{N}\bar{N}  + \frac{1}{2}y TTH_1 +y'T\bar{F}H_2^{\dagger} + y''\bar{N}\bar{F}H_2  +h.c. \nonumber\\
     &+ \lambda_{\rm PQ}\frac{H^\dagger_1 H_2\Phi_{\rm PQ}^5}{M_{\rm P}^3} + g\frac{\Phi_{PQ}^{15}}{M_P^{11}} + ...
\end{align}

It is now clear that the three neutrinos have tiny Majorana masses via the seesaw mechanism \cite{Minkowski:1977sc, Yanagida:1979as, Ramond:1979ijt} after the electroweak symmetry breaking ($\expval{H_{1,2}}=v_{1,2}$). And we have a successful leptogenesis to produce the universe's baryon asymmetry \cite{Buchmuller:2005eh} when the Yukawa coupling $y''$ have CP-violating complex phases.

As shown in the previous section we have a global symmetry $U(1)_{PQ}$, which is explicitly broken only by the higher dimensional operator in (\ref{eq:PQbreaking}). It is indeed surprising the global symmetry arises naturally from the discrete gauge symmetry $\mathbb{Z}_4\times \mathbb{Z}_3\times \mathbb{Z}_5$. Since the global $U(1)_{PQ}$ has the QCD anomaly, the strong CP problem is solved dynamically \cite{Peccei:1977ur}. For the charge assignment of the global $U(1)$ symmetry for each field see Table \ref{tab:Z4xZ3xZ5}.

The PQ symmetry is spontaneously broken by the condensation of the $\Phi_{PQ}$ with $\expval{\Phi_{PQ}}=V_{PQ}$ which generates the massless axion. The axion solves the strong CP problem in QCD \cite{Peccei:1977ur}. The axion receives a small mass $m_a$ from the QCD instanton integration as
\begin{align}\label{eq:coupling}
   m^2_a\simeq \frac{\sqrt{m_um_d}}{m_u+m_d} N_D \frac{m_\pi f_\pi}{f_a}.
\end{align}
Where the decay constant $f_a$ of the axion is given by $f_a=V_{PQ}/N_D$. The other constants are the domain-wall number ($N_D=15$)\footnote{The domain wall number is model dependent. In particular it depends on the number of colored fermions and their PQ charges. For the previous case \cite{Sheng:2025sou} in particular it is $N_D =12$.} and $m_u \simeq 2.16$ MeV, $m_d \simeq 4.67$ MeV and $m_{\pi} \simeq 140$ MeV and $f_\pi = 92$ MeV, which are the masses of the up and down quarks, and the pion mass and the pion decay constant.

The axion gets an extra mass from the allowed higher dimensional operator in (\ref{eq:PQbreaking}). If these corrections to the axion mass are large we lose the solution to the strong CP problem. We find there is no problem if the coupling $g <\mathcal{O}(1)$ and the axion can be the solution to the strong CP problem. We call it a "high-quality axion". Here, we have assumed $V_{PQ}\simeq 0.5\times 10^{13}$GeV. The axion can be the DM in the universe see \cite{Kawasaki:2026jen} for details of its phenomenology. The axion density in the present universe can be derived by means of the misalignment mechanism \cite{Marsh:2015xka,Abbott:1982af,Preskill:1982cy,Dine:1982ah}, which is given by, in terms of the misalignment angle $\theta_a$, 
\begin{align}\label{eq:axion density}
   \Omega_ah^2 \simeq 0.12\qty(\frac{f_a}{ 10^{12} \ \text{GeV}})^{7/6}\qty(\frac{\theta_a}{0.8})^2 ,
\end{align}

The axion has quantum fluctuations during the inflation, $\delta a\simeq H_{inf}/2\pi f_a$. The constraints on the iso curvature fluctuation gives a very strong upper bound on the inflation scale $H_{inf} \le 10^{8}$ GeV. Such a low scale inflation is obtained in so called hilltop inflation that can be realized easily in the present framework \cite{Kawasaki:2023zpd}. On the other hand, if the inflation scale is much higher as $H_{inf}\simeq 10^{13}$ GeV, the DM axion hypothesis has a serious iso curvature problem.
However, it has been, recently, shown the problem can be solved by the improved Linde mechanism. See \cite{Kawasaki:2026jen} for details.

\section{Discussion and conclusions}

In this letter, we propose the no-scale gravity where the $\phi=0$ point is removed. The $\phi$ is the BD type boson field. This is possible since the origin is the fixed point of the scale transformation, $\phi\xrightarrow{} k\phi$ and $g_{\mu\nu} \xrightarrow{} k^{-2}g_{\mu\nu} $ with $k=$constant. However, we have a mathematical problem, since two regions, $\phi <0$ and $\phi>0$ are disconnected. Thus, we introduce $\mathbb{Z}_4$ gauge symmetry to identify the two disconnected regions, where the $\phi$ have a charge $+2$.

With the above setup we show that the right-handed neutrinos naturally have masses, $\lambda_N \phi\bar N\bar N$ assuming the $\mathbb{Z}_4$ charge of the right-handed neutrinos $\bar N$ to be $+1$. However, we have discrete gauge DF anomalies \cite{Dai:1994kq}. To cancel the anomaly we give non-trivial charges under $\mathbb{Z}_4$ to the SM sector. However, the SM particles can be neutral if the mass term of the right-handed neutrinos is given by $\sqrt{\phi^2}\bar N\bar N$ since $\phi^2$ is invariant under the gauge symmetry. In this case, the right-handed neutrinos have a zero charge of the $\mathbb{Z}_4$ and the discrete symmetry $\mathbb{Z}_5$ in Table \ref{tab:Z4xZ3xZ5} can be replaced by almost any integers  like $N=4,7,9...$ to suppress sufficiently unwanted $U(1)_{PQ}$-breaking operators. And as a result the PQ breaking scale can be taken larger values and the domain-wall number depends on the unfixed $N$. Therefore, the PQ-breaking scale and the domain-wall number in future experiments may be able to shed some light on the reliability of No-Scale Gravity.

If one excludes the possibility of having terms like $\sim|\phi|$ then the discrete gauge symmetry give tighter constraints on the theory. Besides restricting the charges and consequently the PQ breaking scale and domain-wall number, it may also restrict the degree of divergence of loop diagrams. It is natural to assume all dimensional parameters are given by integer powers of the boson field $\phi$. If it is the case, all divergences in quantum loops have also integer powers of the $\phi$. Since the divergences must be $\mathbb{Z}_4$ gauge invariant all odd-number divergences must be vanish. Thus, there is no tadpole divergence, for instance, in our theory. Here we have assumed the cutoff scale $\sim \sqrt{\xi}\phi$ so that the scale invariance is maintained in the original theory.

\medskip

\textbf{\textit{Acknowledgements.}}

We thank K. Mukaida and M. Hong for discussion on the convention of relevant fields. This work was supported in part by JSPS KAKENHI No.  JP24H02244 (T.~T.~Y.), JSPS KAKENHI Grant No. JP26K17133 (A.~H.), and the
CD3 Google Seed grant (A.~H.).  A. H. and T. T. Y.  were also supported by World Premier International Research Center Initiative (WPI Initiative), MEXT, Japan. L.~dS.~P. was supported by “Fundação de Amparo à Pesquisa do Estado de São Paulo” (FAPESP) under contracts 2025/24242-0 and 2024/16149-8.

\bibliography{references}

\clearpage
\appendix
\setcounter{equation}{0}
\setcounter{table}{0}
\setcounter{figure}{0}
\renewcommand{\thetable}{A\Roman{table}}
\renewcommand{\thefigure}{A\arabic{figure}}
\renewcommand{\theequation}{A\arabic{equation}}

 \section{Discrete symmetry Anomaly}\label{app:Dai-Freed_Anomaly}


Here we give a brief explanation on how to calculate the anomalies of discrete symmetries and justify the choice of charges done in the paper \cite{Csaki:1997aw}. The basic idea is to use the usual anomaly calculation procedure by enlarging the $\mathbb{Z}_n$ discrete symmetry to a $U(1)_X$ symmetry such that if $U(1)_X$ is non anomalous so is $\mathbb{Z}_n$. Since we don't really want the $U(1)_X$ symmetry we may introduce an artificial scalar field $\rho$ with charge $\pm n$ to break $U(1)_X\to \mathbb{Z}_n$. Here, consider for instance the charges given in Table \ref{tab:Z4xZ3xZ5}. For the $\mathbb{Z}_3$ anomaly, we must calculate the triangle diagrams :
\begin{enumerate}
    \item $U(1)_X\times \text{SM}\times \text{SM}$: This diagram is proportional to $N_G\cdot X\cdot \tr[\lambda_a\lambda_b]$, where $\lambda^a$ are the Gell-mann matrices, $X$ is the charge under $U(1)_X$ and $N_{G}$ is the multiplicity of representation of the group $G$ which is not being traced over. Note that $-2$ mod 3 $= 1$ and hence each generation has charges $(1,1,-2)$ under $U(1)_X$. Therefore each diagram contributes with coefficients $1+1-2 =0$ hence no contribution to the anomaly.
    \item $U(1)_X\times U(1)_X\times \text{SM}$: This is proportional to $X^2\cdot\tr{\lambda^a}$, since the Gell-Mann matrices are traceless  we have 0 contribution.
    \item $U(1)_X \times U(1)_X \times U(1)_X $: This is the dangerous diagram, since we have that $T$ is in a \textbf{10} representation and $\bar{F}$ in $\mathbf{\bar{5}}$, they only contribute with an overall $15$ coefficient and the cancellation must come from the sum of the charges cubed, meaning the diagrams are proportional to $X^3$, hence the overall factor of the triangle diagram is $(1+1-2^3)\cdot 15 = -90$.
\end{enumerate}

We then have that to cancel the $U(1)_X$ anomaly, we must introduce another pair of fermions, say $\chi'$ and $\chi$ with charges $+1$ and $+2$, respectively. This gives a contribution of $1+2^3 = 9 $, if we then have $10$ pairs of them, it gives an overall anomaly factor of $90$. It seems then that we now have 10 new fermions in the spectrum. However, because of the charges, we can write   $U(1)_X$ invariant mass terms for them, given that the scalar field $\rho$ has  a charge -3 we could write $\sim \rho \chi \chi'$ if $\expval{\rho}\sim M_{\rm p}$ then they disappear from the low-energy spectrum of the theory. 

By similar arguments, one can show that by introducing the right-handed neutrinos the $\mathbb{Z}_4$ is non-anomalous, same conclusion can be drawn for the $\mathbb{Z}_5$ as both can be seen as coming from a $U(1)_{B-L}$. A generic $\mathbb{Z}_n\subset U(1)_{B-L}$ has the non anomalous combination of charges

\begin{table}[h]
    \centering
    \setlength{\tabcolsep}{12pt}
    \begin{tabular}{cccccc}
        \hline\\[-2.0ex]
         & $T$ & $\bar{F}$ & $\bar{N}$ \\
        \hline\hline
        $\mathbb{Z}_n$ & 1 & n-3 & 5-n  \\ 
        $U(1)_{B-L}$ & 1 & -3 & 5\\
        \hline
    \end{tabular}
    \caption{Charges of the fermions under $\mathbb{Z}_n$ and $U(1)_{B-L}$ which is non-anomalous.}
    \label{tab:Zn_non_anomalous}
\end{table}

This is consistent with the choice made in the text and hence all the three discrete gauge groups are non-anomalous as should be.

\end{document}